\documentclass[default]{sn-jnl}
\usepackage{etoolbox}
\usepackage{setspace}
\usepackage{bm}
\usepackage{float}
\usepackage{amssymb}
\usepackage{braket}
\usepackage{placeins}
\usepackage{lineno}
\usepackage{bm}
\usepackage{amssymb}
\usepackage{braket}
\usepackage{placeins}
\usepackage{graphicx}%
\usepackage{multirow}%
\usepackage{amsmath,amssymb,amsfonts}%
\usepackage{amsthm}%
\usepackage{mathrsfs}%
\usepackage[title]{appendix}%
\usepackage{xcolor}%
\usepackage{textcomp}%
\usepackage{manyfoot}%
\usepackage{booktabs}%
\usepackage{algorithm}%
\usepackage{algorithmicx}%
\usepackage{algpseudocode}%
\usepackage{listings}%
\usepackage{commath}
\usepackage{natbib}
\setcitestyle{round}
\makeatletter
\renewcommand{\@biblabel}[1]{#1.}
\makeatother
\usepackage{etoolbox}
\makeatletter
\patchcmd{\ps@headings}
{\hbox to \hsize{\hfill Springer Nature 2021 \LaTeX\ template\hfill}}
{\hbox to \hsize{\hfill  \hfill}}
{}
{}
\patchcmd{\ps@headings}
{\hbox to \hsize{\hfill Springer Nature 2021 \LaTeX\ template\hfill}}
{\hbox to \hsize{\hfill  \hfill}}
{}
{}
\patchcmd{\ps@titlepage}
{\hbox to \hsize{\hfill Springer Nature 2021 \LaTeX\ template\hfill}}
{\hbox to \hsize{\hfill  \hfill}}
{}
{}
\makeatother

\jyear{2023}%

\theoremstyle{thmstyletwo}%

\theoremstyle{thmstylethree}%
\unnumbered

\begin{document}

\title[SSAI-3D: System- and Sample-agnostic Isotropic 3D Microscopy]{System- and Sample-agnostic Isotropic 3D Microscopy by Weakly Physics-informed, Domain-shift-resistant Axial Deblurring}
\author[1,2]{\large Jiashu Han}
\equalcont{\small These authors contributed equally to this work.}
\author[1,2]{\large Kunzan Liu}
\equalcont{\small These authors contributed equally to this work.}
\author[3]{\large Keith B. Isaacson}
\author[1]{\large Kristina Monakhova}
\author[4,5]{\large Linda G. Griffith}
\author*[1,2]{\large Sixian You}\email{\small sixian@mit.edu}

\affil[1]{\small \centering{Research Laboratory of Electronics, MIT}\vspace{1mm}}
\affil[2]{\small \centering{Department of Electrical Engineering and Computer Science, MIT}\vspace{1mm}}
\affil[3]{\small \centering{Newton-Wellesley Hospital, Mass General Brigham}\vspace{1mm}}
\affil[4]{\small \centering{Department of Biological Engineering, MIT}\vspace{1mm}}
\affil[5]{\small \centering{Department of Mechanical Engineering, MIT}\vspace{1mm}}

\abstract{
Three-dimensional (3D) subcellular imaging is essential for biomedical research, but the diffraction limit of optical microscopy compromises axial resolution, hindering accurate 3D structural analysis. This challenge is particularly pronounced in label-free imaging of thick, heterogeneous tissues, where assumptions about data distribution  (e.g. sparsity, label-specific distribution, and lateral-axial similarity) and system priors (e.g. independent and identically distributed (i.i.d.) noise and linear shift-invariant (LSI) point-spread functions (PSFs)) are often invalid. Here, we introduce SSAI-3D, a weakly physics-informed, domain-shift-resistant framework for robust isotropic 3D imaging. SSAI-3D enables robust axial deblurring by generating a PSF-flexible, noise-resilient, sample-informed training dataset and sparsely fine-tuning a large pre-trained blind deblurring network. SSAI-3D was applied to label-free nonlinear imaging of living organoids, freshly excised human endometrium tissue, and mouse whisker pads, and further validated in publicly available ground-truth-paired experimental datasets of 3D heterogeneous biological tissues with unknown blurring and noise across different microscopy systems.

}

\maketitle

\newpage
\section{Introduction}\label{sec:main}

Three-dimensional (3D) sub-cellular imaging is highly desirable in biomedical research, as it reveals the intricate spatial organization of organelles, cells, and biological networks. Achieving high-resolution sub-cellular imaging across all three dimensions is crucial for accurately understanding complex biological processes, such as neural circuits, disease pathogenesis, and cellular responses to drugs~\cite{gobel2007imaging,wang2018three,cruz2019neutrophil,olivier2010cell,pulous2022cerebrospinal,zhao2023two}. However, the fundamental diffraction limit of optical microscopy results in axial resolution that is 2–5 times worse than lateral resolution. This resolution disparity severely hinders isotropic 3D imaging, limiting our ability to faithfully represent the true 3D structure of biological samples and fully realize the potential of 3D imaging in scientific research~\cite{boden2021volumetric,verveer2007high,zipfel2003nonlinear}.

While recent advances in microscopy theory and instrumentation enable isotropic 3D imaging with certain specialized systems, their reliance on specific hardware, contrast mechanisms, and/or sample preparation limits widespread adoption, particularly for thick, living, and intact biosystems~\cite{planchon2011rapid,aquino2011two,jia2014isotropic,chhetri2015whole,wu2021multiview,zhao2022isotropic,dean2022isotropic,li2023three}. To make isotropic imaging more accessible across different microscopy modalities, algorithmic approaches seek to address the inverse problem of axial deblurring. In recent years, deconvolution methods based on deep neural networks (DNN), given their effective and flexible learning of data priors, have shown unprecedented performance compared to classical deconvolution methods, such as Richardson-Lucy and fast iterative shrinkage thresholding algorithm (FISTA), on two-dimensional (2D) deconvolution~\cite{ren2020neural,khan2020flatnet,whang2022deblurring,yanny2022deep,kohli2022ring}. However, axial resolution restoration is significantly more challenging since acquiring isotropic 3D imaging data as ground truth is impractical for most microscopes, especially in living tissues, making supervised learning methods unsuitable for general isotropic 3D imaging. Despite the existence of a few systems capable of collecting paired isotropic 3D imaging data and generously sharing it~\cite{wu2021multiview,voigt2019mesospim}, the potential for pre-training a network and generalizing to diverse microscopes and samples is hampered by the significant domain shifts between training and target domains inherent in biomedical imaging~\cite{moor2023foundation,ma2024pretraining}.

Domain shift is particularly pronounced in scientific imaging compared to general computer vision~\cite{hendrycks2016baseline,hendrycks2021many,andreassen2022evolution}, due to the high heterogeneity of microscopy modalities, system properties, sample variation across species and tissues, and the exploratory nature (unseen anomaly detection) of microscopy data.
To address these challenges and enable general isotropic 3D imaging, unsupervised generative adversarial network (GAN)-based methods have been proposed to enable deblurring without matched axial image pairs~\cite{li2021unsupervised,park2022deep,ning2023deep}.
However, the inherent instability of saddle-point optimization in GAN-based methods leaves them vulnerable to network collapse, potentially generating hallucinations and inconsistencies~\cite{yadav2018stabilizing,cohen2018distribution,de2021deep}.
Supervised methods leverage similarities between lateral and axial data distributions and employ explicit blurring forward models to achieve high-fidelity isotropic imaging without paired in-distribution data~\cite{weigert2018content,chen2021three}. However, the practical applications of single-stack isotropic recovery using supervised methods are limited for two primary reasons.
First, these methods rely on an explicit point-spread function (PSF) model that is assumed to be precisely matched and consistent across the entire imaging volume. In reality, factors such as field curvature, misalignment, and tissue scattering contribute to optical aberrations, resulting in spatially varying PSFs that are sample- and system-dependent, making real-time calibration difficult. This modeling challenge is further exacerbated by noise, especially in low-light conditions~\cite{guo2020rapid}.
Second, self-supervised methods require assumptions of lateral-axial similarity within single-stack data to learn axial deblurring from synthetic lateral deblurring. While this assumption holds for certain cell and tissue structures (e.g., neurons in the cortex and sinusoids in the liver, as shown in prior axial deblurring works~\cite{weigert2018content,park2022deep,ning2023deep}), it breaks down in tissues with high directionality, polarity, or anisotropy, which are common in biological systems such as developing embryos, epithelial glands, and collagen fibers in the extracellular matrix~\cite{guo2023deep}.

These challenges are particularly pronounced in label-free imaging of thick, heterogeneous tissues~\cite{liu2024deep}, where assumptions about data distribution  (e.g. sparsity, label-specific distribution, and lateral-axial similarity) and system priors (e.g. independent and identically distributed (i.i.d.) noise, and linear shift-invariant (LSI) PSFs) are often invalid. To overcome these challenges, we propose SSAI-3D (System- and Sample-agnostic Isotropic 3D Microscopy), a weakly physics-informed, domain-shift-resistant axial deblurring framework for robust isotropic recovery across diverse systems and samples.
First, we formulate isotropic recovery as a semi-blind deblurring problem and leverage a pre-trained blind deblurring network developed for natural scene images as the initial network. Then, we create a synthetic training dataset by blurring denoised lateral images with various PSFs of different sizes and orientations to adapt the deblurring network to the current 3D imaging stack. This allows us to train the network in a self-supervised manner, eliminating the need for pixel-wise co-registered data pairs. We demonstrate that the combination of a denoised and PSF-varying training dataset, coupled with leveraging existing knowledge from the experimentally pre-trained blind deblurring network, significantly alleviates susceptibility to spatially varying 3D blurring kernels and noise.

Secondly, we sparsely fine-tune the large pre-trained deblurring network using this synthetic dataset. A surgeon network is employed to identify which layers are most critical for adapting to microscopic images, while other layers remain frozen to preserve the original deblurring capabilities learned from the much larger dataset of natural scene image pairs.
This approach leverages prior knowledge of blind deblurring while efficiently adapting to microscopy data and minimizing assumptions about lateral-axial similarity, thus reducing overfitting and computational costs commonly associated with microscopy deconvolution problems.
We demonstrate high-fidelity single-stack axial deblurring performance of SSAI-3D on open-source experimental datasets from various microscopy systems: confocal (with near-isotropic validation through hardware modification~\cite{voigt2019mesospim}), light-sheet (with near-isotropic validation through hardware modification~\cite{wu2021multiview}), wide-field, and nonlinear microscope~\cite{you2018intravital,liu2024deep}.
The high-fidelity recovery across diverse microscopy systems (varying in aberrations, noise, and contrasts) and samples (with different degrees of lateral-axial dissimilarity) indicates that SSAI-3D can be a widely accessible and reliable tool for isotropic imaging, facilitating precise analysis of complex 3D biological structures and processes.

\section{Results}\label{sec:results}

\subsection{Isotropic resolution recovery by self-supervised sparsely fine-tuned deblurring}
\label{tuning}

\begin{figure}[t]
\centering
\includegraphics[width=1\textwidth]{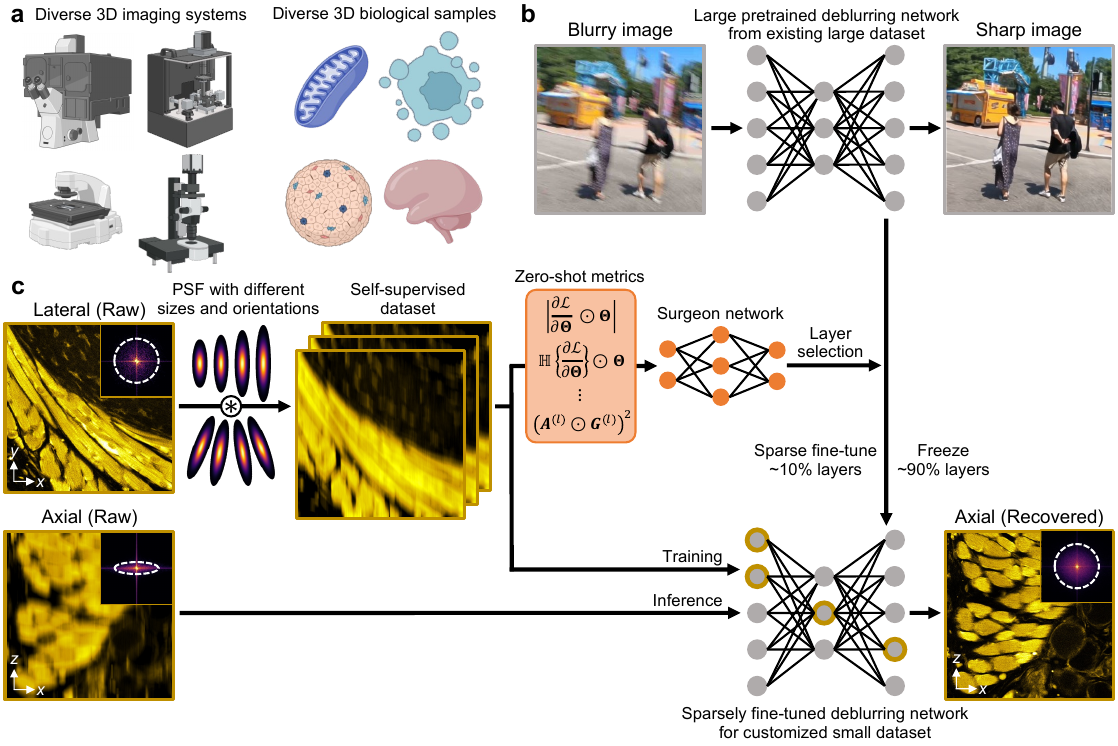}
\caption{\textbf{Principles of SSAI-3D.}
\textbf{a}, SSAI-3D enables robust isotropic resolution recovery across diverse 3D imaging systems (confocal, light-sheet, wide-field, and nonlinear) and diverse 3D biological samples (organelles, cells, tissues, and organs).
\textbf{b}, The deblurring network is initialized with a large pre-trained network for blind deconvolution on a large dataset of natural image pairs.
\textbf{c}, Starting with a single microscopy-specific image stack where the axial resolution is worse than the lateral resolution, lateral images are blurred with a series of PSFs of different sizes and orientations to generate a self-supervised dataset. Then, generating zero-shot metrics using $\sim$1\% of this dataset, a surgeon network is employed to select the critical layers to fine-tune in the large pre-trained deblurring network.
Only $\sim$10\% layers are selected and sparsely fine-tuned according to the generated self-supervised dataset. Given unseen axial images, the fine-tuned deblurring network predicts deblurred images with isotropic resolution. Insets in the images represent the corresponding Fourier spectrums.
}
\label{fig1}
\end{figure}

The SSAI-3D framework for single-stack isotropic resolution recovery consists of two steps (Fig.~\ref{fig1}c). First, a self-supervised training data is generated by applying a series of PSFs with varying sizes and orientations to denoised lateral images (see Methods for details). Unlike methods that rely on an exact PSF model and thus have strong physics-based assumptions about the image generation forward model, these sets of weakly-physics-based synthetic pairs emphasize the general blurring action of the optical microscope rather than the precise forward model, thereby increasing the network's robustness against unknown optical aberrations and noise.

Secondly, we leverage knowledge learned from natural scene deblurring tasks via a large pre-trained deblurring network (Fig. \ref{fig1}b and Supplementary Fig. \ref{fig_network}a). We then sparsely fine-tune it using the created self-supervised dataset, selecting and fine-tuning only $\sim$10\% of the layers while freezing the rest. 
Layer selection is guided by a surgeon network, which evaluates each layer's contribution to the adaptation task using zero-shot metrics, including statistics related to the activation, gradient, loss function, and pre-trained weights (see Methods for details and Supplementary  Fig.~\ref{fig_network}b for architecture).
This surgeon network mitigates the stochastic instability of zero-shot metrics, while sparse fine-tuning avoids performance degradation in out-of-distribution data and the computational burden of full fine-tuning~\cite{lee2022surgical}.
Our approach preserves the deblurring capability learned from the larger natural scene dataset, making only necessary adaptations to the microscopy image stack. This results in a computationally and data-efficient process that is less susceptible to overfitting to the lateral data distribution or potential mismatches in the forward model.
Finally, with these two steps, in the inference stage, the axial images are sent to the adapted customized deblurring network that predicts deblurred images with isotropic resolution.
Without strong assumptions about the imaging system or sample data distribution, SSAI-3D is anticipated to be a widely accessible and robust tool for isotropic 3D imaging across diverse microscopy modalities and a broad spectrum of biological applications (Fig.~\ref{fig1}a).

\begin{figure}[!h]
\centering
\includegraphics[width=\textwidth]{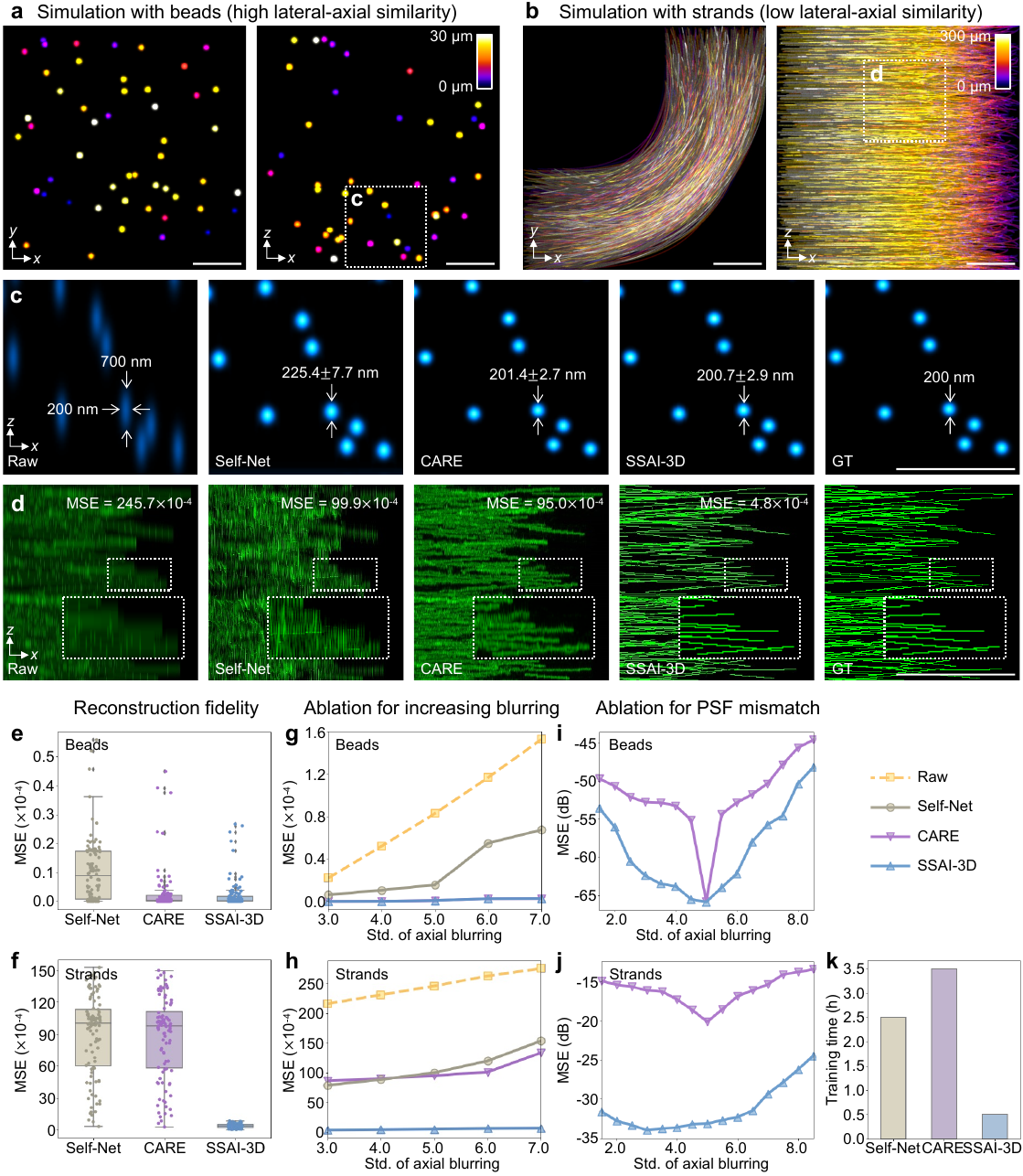}
\caption{\textbf{Performance of SSAI-3D on synthetic samples.}
Simulation of bead (\textbf{a}) and strand structures (\textbf{b}) reveals the performance of SSAI-3D and existing methods (\textbf{c}, \textbf{d}). 
The simulation objects are visualized using depth-coded projections with the colorbar representing the actual depth.
Reconstructed axial resolution (mean$\pm$standard deviation) is labeled in \textbf{c}.
\textbf{e, f}, Characterization of reconstruction fidelity for beads (\textbf{a}) and strands (\textbf{b}). 
\textbf{g, h}, Ablation study on the performance of different levels of blurring for beads (\textbf{a}) and strands (\textbf{b}).
\textbf{i, j}, Robustness of reconstruction against PSF mismatch for beads (\textbf{a}) and strands (\textbf{b}). 
A blurring with a standard deviation of 5 is assumed to be known for CARE.
The standard deviation of the axial PSF on the x-axis of (\textbf{g}--\textbf{j}) is also the times worse it is relative to the lateral PSF.
\textbf{k}, Comparison of the training time.
Scale bars: 6\,\textmu m (\textbf{a}, \textbf{c}); 60\,\textmu m (\textbf{b}, \textbf{d}).}
\label{fig2}
\end{figure}

We first compared SSAI-3D to the state-of-the-art unsupervised (Self-Net~\cite{ning2023deep}) and supervised (CARE~\cite{weigert2018content}) methods using two distinct synthetic objects, assuming a known and LSI PSF across the imaging volume.
The first synthetic object comprised 50 randomly seeded sub-diffraction-limited fluorescent beads within a 30\,\textmu m$\times$30\,\textmu m$\times$30\,\textmu m imaging volume (Fig.~\ref{fig2}a).
Assuming a lateral resolution of 200\,nm and axial resolution of 700\,nm for the imaging system, all three methods demonstrated significant improvement in axial resolution (Fig.~\ref{fig2}c).
While Self-Net exhibited greater variability in bead shape and intensity, both CARE and SSAI-3D achieved reliable axial resolution recovery (201.4$\pm$2.7\,nm and 200.7$\pm$2.9\,nm, respectively).

However, many biological structures do not exhibit perfect lateral-axial similarity like round beads. Therefore, we conducted a second simulation on a more heterogeneous 3D synthetic structure, comprising 5000 directional strands within a 300\,\textmu m$\times$300\,\textmu m$\times$300\,\textmu m imaging volume (Fig.~\ref{fig2}b; see Methods for details).
This denser object with directionality better simulates the complex 3D structures found in real biological samples, such as dendrites, glands, and fibers (neural, elastin, and collagen).
In contrast to the bead simulation, existing methods exhibited significant performance degradation, while SSAI-3D maintained high-fidelity resolution recovery with accurate shape and intensity preservation (Fig.~\ref{fig2}d and Supplementary Fig.~\ref{fig_line}).
Quantitative analysis further confirmed that, even with a known and LSI PSF, existing algorithms significantly deteriorate in performance when applied to directional, 3D asymmetric structures, as a result of their reliance on the assumption of lateral-axial similarity.
In contrast, SSAI-3D produced high-fidelity restorations with low mean square error (MSE) (Fig.~\ref{fig2}e and f) and high structural similarity (SSIM) (Supplementary Fig.~\ref{fig_SSIM}) across both simulations. 
Ablation studies confirmed the consistent axial deblurring fidelity of SSAI-3D across varying levels of blurring (Fig.~\ref{fig2}g and h) and its robustness to PSF mismatch (Fig.~\ref{fig2}i and j; see Methods for details).
In addition, benefiting from sparse fine-tuning with only $\sim$15 million parameters modified, SSAI-3D exhibited a 2.5$\sim$3.5-fold reduction in training time compared to existing methods, thereby enhancing its accessibility and reducing computational requirements (Fig.~\ref{fig2}k).

\subsection{Sparse fine-tuning on large pre-trained model relaxes assumptions on lateral-axial similarity of 3D data}

\begin{figure}[t]
\centering
\includegraphics[width=\textwidth]{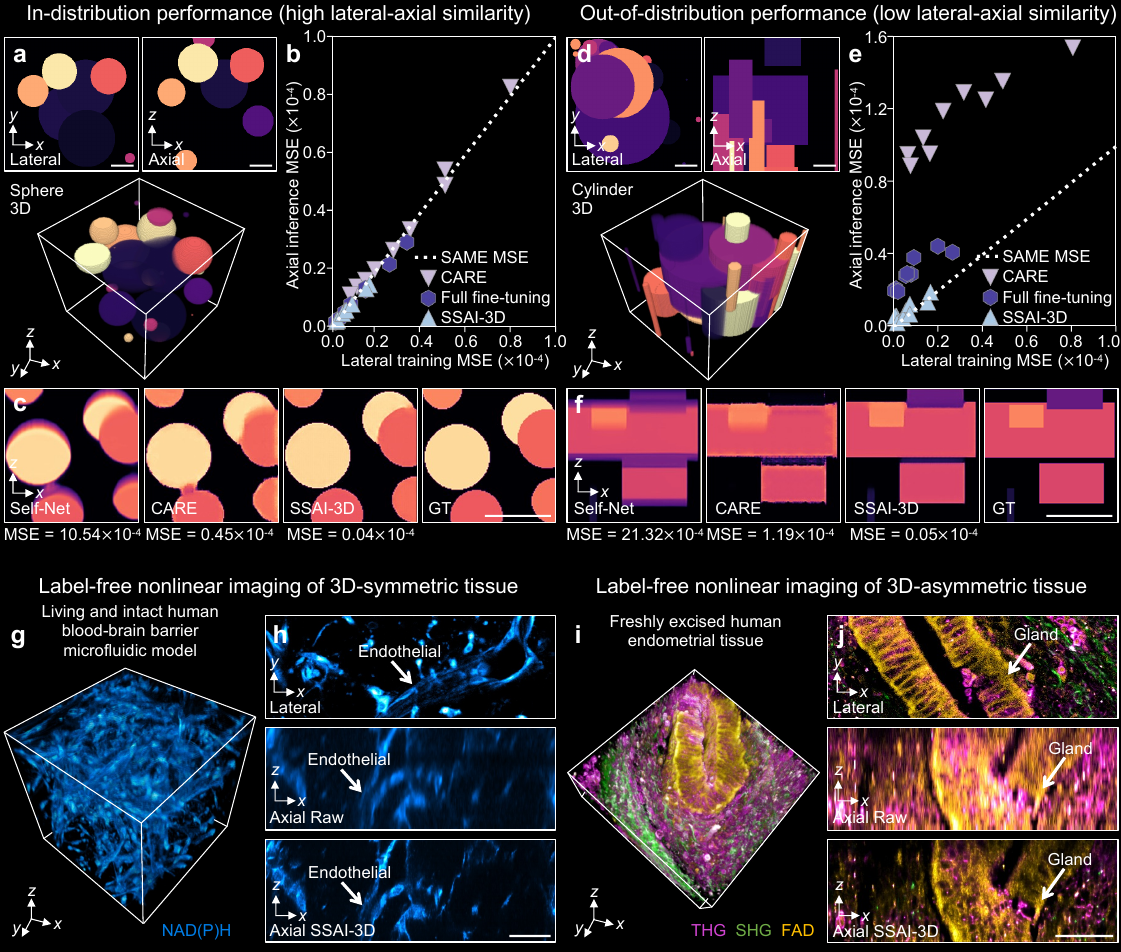}
\caption{\textbf{Effect of lateral-axial similarity of 3D data using simulations and tissues.}
Simulation of sphere (high lateral-axial similarity, \textbf{a}) and cylinder (low lateral-axial similarity, \textbf{d}) reveals the performance of SSAI-3D and existing methods (\textbf{c}, \textbf{f}). 
\textbf{b, e}, Relationship between training MSE on lateral images and inference MSE on axial images indicates robustness of different methods against distribution shifts.
In label-free nonlinear imaging, biological tissues exhibit different levels of lateral-axial similarity (high in living and intact human blood-brain barrier microfludic model (\textbf{g}) and low in freshly excised human endometrial tissue (\textbf{i})).
\textbf{h, j}, Raw lateral and axial images of \textbf{g} and \textbf{i}, as well as restored axial images using SSAI-3D. Arrows: symmetric endothelial cells (\textbf{h}) and polarized epithelial glands (\textbf{j}).
Scale bars: 50\,\textmu m.}
\label{fig3}
\end{figure}

To understand the performance differences between the two data types (beads vs. strands), we further investigated how lateral-axial data similarity influences algorithm performance. We quantified this relationship by plotting the accuracy on the reference lateral distribution against the accuracy on the potentially shifted axial distribution ($n=9$)~\cite{wortsman2022robust}.
In cases of high lateral-axial similarity, existing methods and SSAI-3D aligned with the identity line ($y=x$, Fig.~\ref{fig3}b), indicating successful deblurring when the distribution shift in the test axial datasets is negligible.
However, when lateral-axial similarity is low within the sample, CARE and SSAI-3D with full fine-tuning exhibited a significant gap between training MSE on lateral images and inference MSE on axial images. In contrast, SSAI-3D with sparse fine-tuning demonstrated robustness to distribution shifts, resulting in high-fidelity axial deblurring even in 3D heterogeneous samples~\cite{hendrycks2016baseline,hendrycks2021many,andreassen2022evolution} (Fig.~\ref{fig3}e; see methods and Supplementary Fig. \ref{fig_finetuning} for details in full fine-tuning).

In this experiment, we considered a simulation of imaging spheres and cylinders in 3D, representing cases of high (circles in both lateral and axial planes, Fig.~\ref{fig3}a) and low (circles in lateral, rectangles in axial planes, Fig.~\ref{fig3}d) lateral-axial similarity, respectively. Consistent with the results in Fig.~\ref{fig2}, Self-Net, CARE, and SSAI-3D achieved comparable reconstructions under high lateral-axial similarity (Fig.~\ref{fig3}c). However, in the low similarity case, performance diverged significantly, with SSAI-3D outperforming existing methods (Fig.~\ref{fig3}f). This divergence arises because existing methods heavily rely on lateral-axial similarity, making them susceptible to significant performance decreases when distribution shifts occur in real axial test data. In contrast, the sparse fine-tuning in SSAI-3D preserves deblurring capabilities learned from the natural scene dataset while adapting minimally to the microscopy image stack, resulting in superior axial deblurring performance for 3D heterogeneous samples under lateral-axial distribution shifts.

In real-world biological imaging, lateral-axial similarity can vary significantly depending on the sample's structure and imaging scale. For instance, neurons within a small volume may exhibit high lateral-axial similarity, whereas glands, dendrites, and fibers (neural, elastin, collagen) with high directionality, polarity, or anisotropy typically exhibit low lateral-axial similarity.
We demonstrated this sample-dependent 3D heterogeneity using two thick and living samples imaged with label-free nonlinear imaging~\cite{liu2024deep}. First, we imaged a living human blood-brain barrier microfluidic model where vascularized endothelial cells formed lumen structures with pericytes and astrocytes~\cite{hajal2022engineered}. This 3D self-assembled multicellular model displayed high lateral-axial similarity (Fig.~\ref{fig3}g). Secondly, we imaged a freshly excised human endometrial tissue, where directional collagen growth and the inherent polarity of the epithelial glands resulted in low lateral-axial similarity (Fig.~\ref{fig3}i).
SSAI-3D successfully restored fine structures such as endothelial cells (Fig.~\ref{fig3}h) and epithelial glands (Fig.~\ref{fig3}j) in axial views, even in cases of low lateral-axial similarity. This robustness across samples with varying structural properties highlights the potential of SSAI-3D as a widely applicable tool for 3D biological imaging.

\subsection{Generalizability to varying system imperfections and sample heterogeneity in simulation and experiments}

\begin{figure}[t]
\centering
\includegraphics[width=\textwidth]{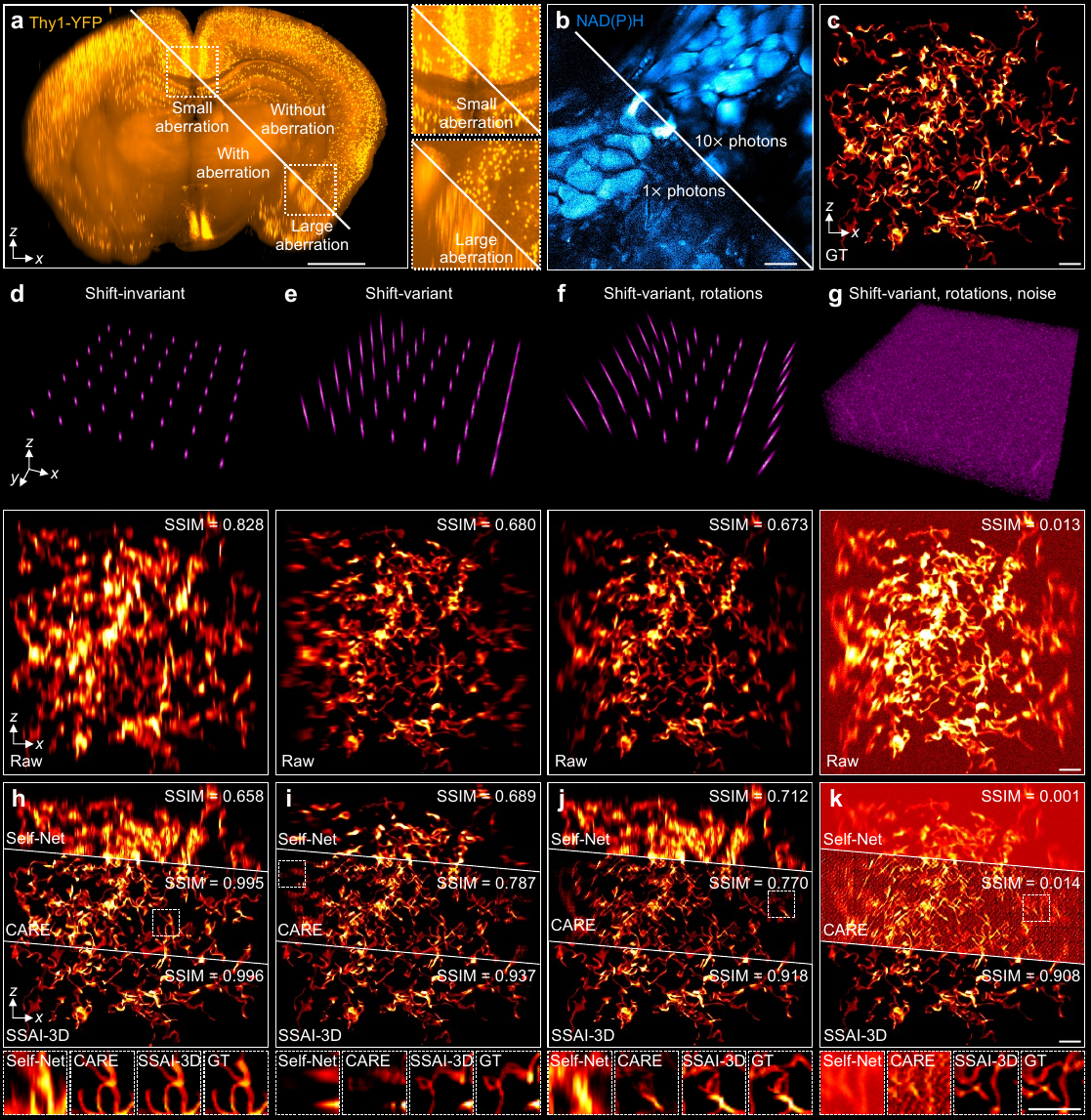}
\caption{
\textbf{Effect of optical aberrations and noise on deblurring performance using simulations.}
\textbf{a}, Example of optical aberrations in real microscopy system (spatially-varying PSF in mesoSPIM \cite{voigt2019mesospim}).
\textbf{b}, Example of noise in real microscopy system (low-light condition in dSLAM \cite{ye2023learned}).
\textbf{c}, Ground truth image for simulation on effect of optical aberrations and noise.
\textbf{d--g}, Four different levels of optical aberrations and noise with the corresponding raw images (\textbf{d}, Shift-invariant PSF (ideal case); \textbf{e}, Shift-variant PSF; \textbf{f}, Shift-variant PSF with rotations along $z$-direction; \textbf{g}, Shift-variant PSF with rotations along $z$-direction and noise across the entire imaging volume).
\textbf{h--k}, Comparison of Self-Net, CARE, and SSAI-3D on deblurring performance using SSIM in scenarios \textbf{d--g}. Bottom: Blow-ups of the co-registered images within the white box.
Scale bars: 2\,mm (\textbf{a}); 25\,\textmu m (\textbf{b}--\textbf{k}).
}
\label{fig4}
\end{figure}

Next, we investigate the robustness of SSAI-3D against various system imperfections compared to existing methods. In practical 3D biomedical imaging, imaging systems exhibit diverse PSFs in shape and size due to varying signal generation mechanisms and customized instrumentation and acquisition parameters. Furthermore, PSFs can be spatially varying across the imaging volume due to factors such as aberration, misalignment, and tissue scattering (Fig.~\ref{fig4}a).  Quantifying the precise PSFs for an arbitrary image stack is infeasible since they are system- and sample-dependent. Additionally, many imaging systems operate under photon-limited conditions due to biophysical and biochemical constraints, such as the need for high imaging speed or to minimize photobleaching, phototoxicity, and tissue heating~\cite{li2023real}. The resulting low signal-to-noise ratio (SNR) images can significantly affect the accuracy of deconvolution methods~\cite{guo2020rapid} (Fig.~\ref{fig4}b). Given these challenges, we demonstrated the robustness of SSAI-3D against a spectrum of system imperfections, including spatially varying PSFs and noise, by leveraging a self-supervised dataset generated from PSFs with diverse sizes and orientations.

We simulated 5000 randomly oriented tubular objects (high lateral-axial similarity) within a 300\,\textmu m$\times$300\,\textmu m$\times$300\,\textmu m imaging volume (Fig.~\ref{fig4}c).
To evaluate axial deblurring performance, we initially applied a shift-invariant axial blur, assumed to be known for CARE (Fig.~\ref{fig4}d).
In this ideal scenario, Self-Net, CARE, and SSAI-3D all achieved high-quality resolution recovery with low MSE (Fig.~\ref{fig4}h).
However, in practice, obtaining an accurate PSF profile for the system is often challenging, and PSF mismatch can degrade the performance of CARE (Fig.~\ref{fig2}i and j).
Next, we progressively introduced system imperfections. Fig.~\ref{fig4}e and f incorporated variations in the extent and orientation of the axial blur. The non-uniform PSF profiles, combined with asymmetry between lateral and axial aberrations, led to decreased accuracy for Self-Net and CARE, particularly in regions with pronounced aberrations (Fig.~\ref{fig4}i and j).
To incorporate the effect of noise, we introduced both Gaussian and Poisson noise to the images after artificial axial blurring (Fig.~\ref{fig4}g). To mitigate noise corruption and amplification during the forward blurring and inverse deblurring processes, SSAI-3D incorporates a denoising step prior to self-supervised dataset generation and the deblurring model, enhancing its robustness against noise in low-light conditions (Fig.~\ref{fig4}k and Supplementary Fig. \ref{figS_denoise}).

\begin{figure}[t]
\centering
\includegraphics[width=\textwidth]{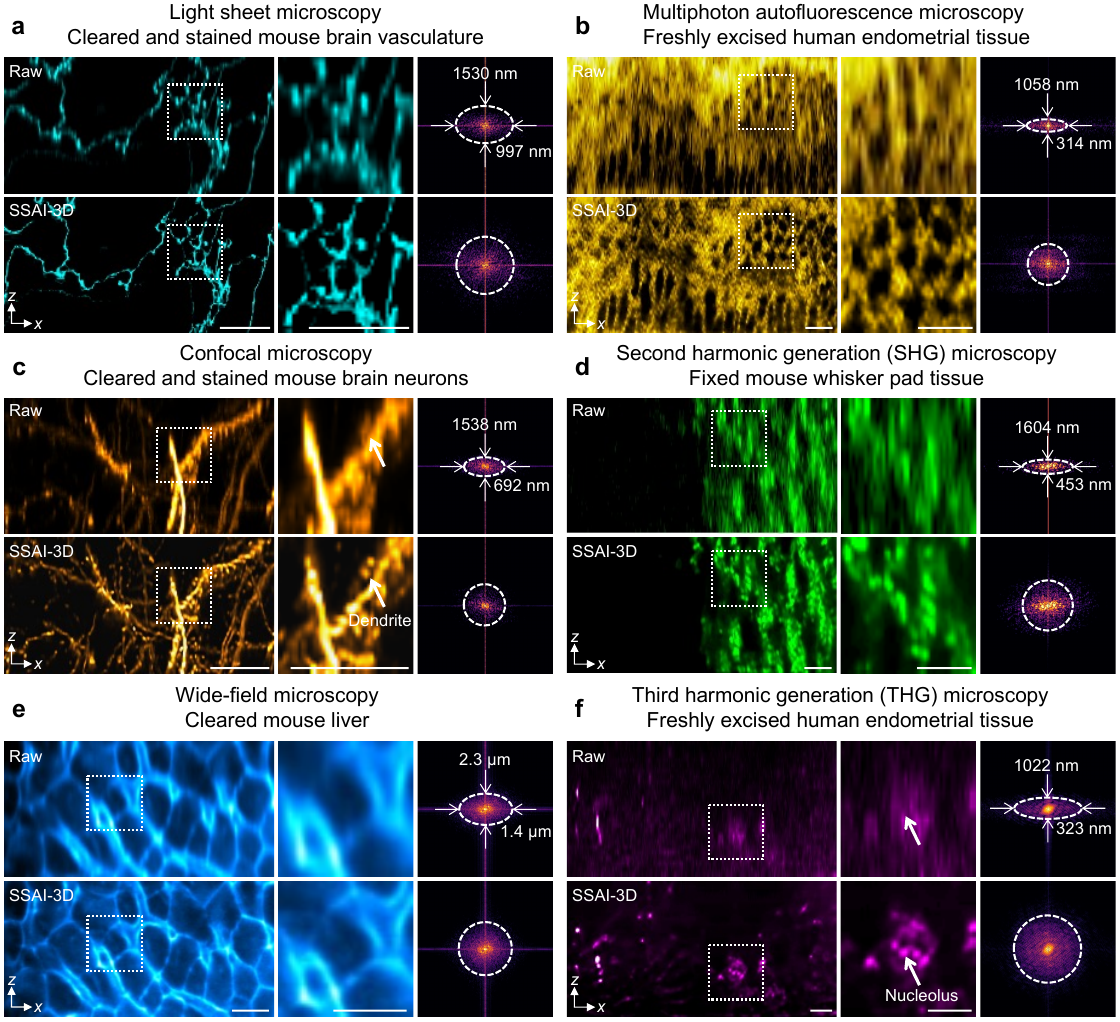}
\caption{\textbf{SSAI-3D allows isotropic resolution recovery across different imaging systems and samples.}
Each example includes the raw and restored images (left), along with their respective blow-ups (highlighted in the white boxes, middle) and Fourier spectrums (right, labeled numbers represent the lateral and axial resolutions).
\textbf{a}, Cleared and stained mouse brain vasculature from light-sheet microscopy.
\textbf{b}, Freshly excised human endometrial tissue from multiphoton autofluorescence microscopy.
\textbf{c}, Cleared and stained mouse brain neurons from confocal microscopy.
\textbf{d}, Fixed mouse whisker follicles from second harmonic generation (SHG) microscopy.
\textbf{e}, Cleared mouse liver from wide-field microscopy.
\textbf{f}, Freshly excised human endometrial tissue from third harmonic generation (THG) microscopy.
Scale bars: 20\,\textmu m.}
\label{fig5}
\end{figure}

Next, we investigate the performance of SSAI-3D on real-life biological data across various microscopy modalities, system properties, and sample types. As demonstrated in Fig.~\ref{fig5}, SSAI-3D achieved consistent isotropic resolution recovery across a broad range of samples (living/fixed, human/animal, varying degrees of lateral-axial similarity) and imaging systems (diverse contrast mechanisms, commercial/custom-built, low/high resolution, low/high SNR). 3D images with varying degrees of anisotropy, acquired using light-sheet, confocal, wide-field, and nonlinear microscopy, all exhibited substantial axial resolution enhancement after restoration with SSAI-3D within 0.5 GPU hours of training (see Supplementary Table~\ref{summary_table} and Methods for details). Fourier spectrum analysis quantified this improvement, revealing previously indistinguishable fine structures in low-resolution raw images, such as dendrites in mouse brain neurons (Fig.~\ref{fig5}c) and nucleoli in human endometrial epithelial cells (Fig.~\ref{fig5}f). Notably, the surgeon network selected different layers for fine-tuning in each imaging stack (Supplementary Fig.~\ref{figS_contribution}). These results highlight the potential of SSAI-3D as a generalizable and accessible tool for 3D imaging, enhancing axial resolution to facilitate the study of 3D tissue architecture and subcellular features.

\subsection*{Validation of SSAI-3D with hardware-corrected isotropic imaging}

\begin{figure}[p]
\centering
\includegraphics[width=\textwidth]{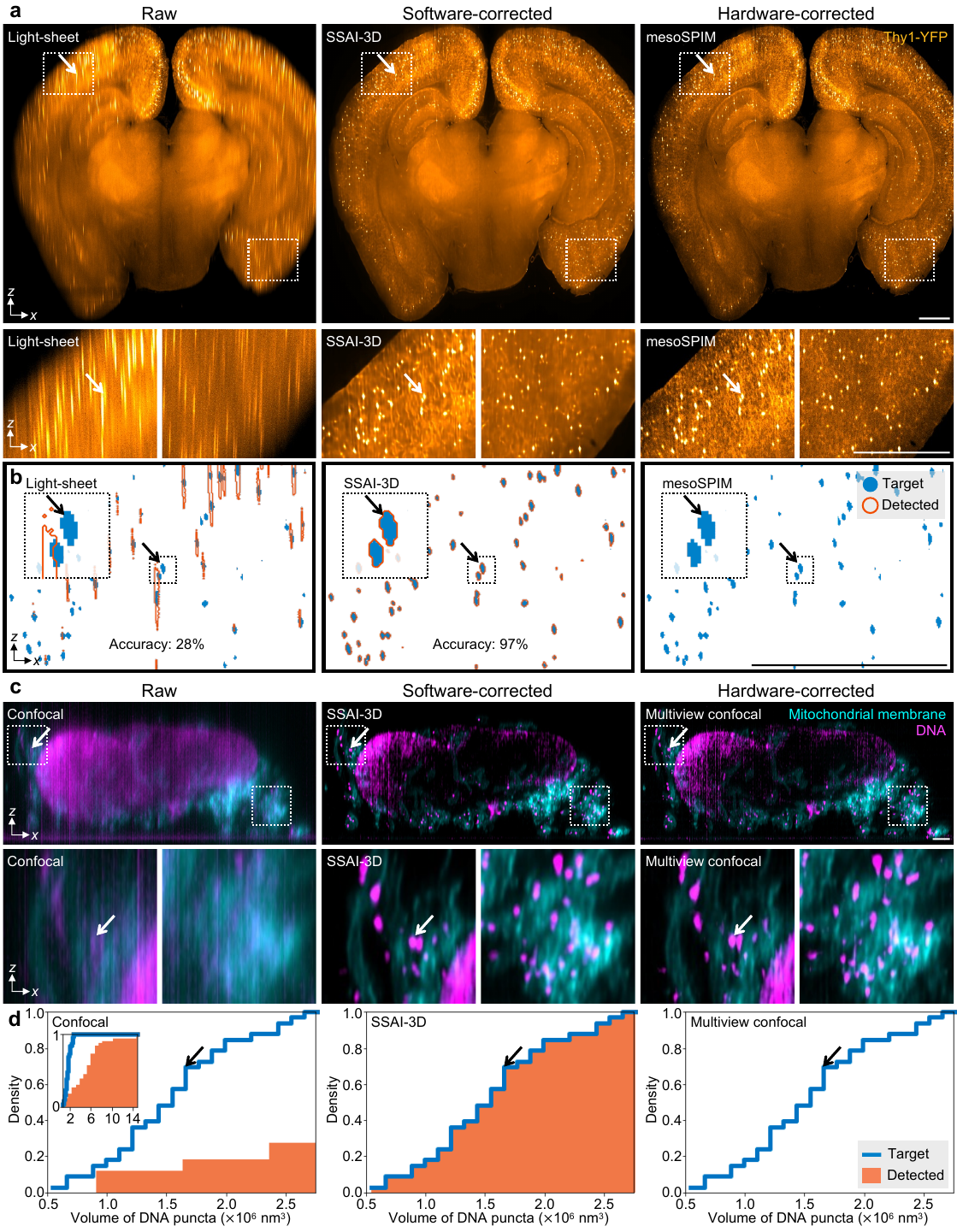}
\caption{\textbf{Validated isotropic resolution recovery of SSAI-3D facilitates downstream biological analysis.}
\textbf{a}, Comparison of software-corrected (SSAI-3D) and hardware-corrected (mesoSPIM) axial image of mouse brain from light-sheet microscopy.
\textbf{b}, Neuron detection accuracy using raw, software-corrected, and hardware-corrected images.
Arrows in \textbf{a} and \textbf{b} are pointing to a same neuron.
\textbf{c}, Comparison of software-corrected (SSAI-3D) and hardware-corrected (multiview confocal) axial image of mitochondria from confocal microscopy.
\textbf{d}, Statistics of the volume of DNA puncta in mitochondria. Inset: overall statistics of raw data.
Arrows in \textbf{c} represent the same DNA puncta with the volume marked in \textbf{d}.
Scale bars: 1\,mm (\textbf{a}, \textbf{b}); 1\,\textmu m (\textbf{c}, \textbf{d}).}
\label{fig6}
\end{figure}

To experimentally validate the high-fidelity axial deblurring of SSAI-3D and demonstrate its robustness to high sample heterogeneity and unknown system imperfections for downstream biological analysis, we presented two comparative studies against near-isotropic imaging references obtained via hardware advancements.
First, we assessed SSAI-3D's isotropic resolution recovery in light-sheet microscopy using co-registered raw images and a near-isotropic reference acquired by mesoscale selective plane-illumination microscopy (mesoSPIM)~\cite{voigt2019mesospim} (Fig.~\ref{fig6}a). A mouse brain was cleared, stained, and imaged with mesoSPIM to visualize the vasoactive intestinal peptide (VIP)-expressing interneurons~\cite{pi2013cortical}. At the mesoscale, the mouse brain exhibited high lateral-axial dissimilarity, and system aberrations varied significantly across the large field of view (FOV) (Fig.~\ref{fig4}a).
By comparing against the near-isotropic hardware-corrected images, SSAI-3D demonstrated high-fidelity 3D isotropic resolution recovery despite the high lateral-axial dissimilarity and the spatially varying PSFs, which degraded the performance of existing methods (please refer to Supplementary Fig.~\ref{fig_ASLM} for method comparison). 
To evaluate the effect of SSAI-3D on downstream biological analysis, we quantified the accuracy of neuron extraction and observed the segmentation accuracy of SSAI-3D closely matched that of the near-isotropic hardware-corrected mesoSPIM dataset (Fig. \ref{fig6}b; see Methods for detection algorithms).

To further evaluate generalizability across diverse system imperfections and sample geometries, we assessed SSAI-3D on a multi-modal confocal microscopy dataset, in addition to the light-sheet microscopy data. This dataset comprised co-registered raw images and a near-isotropic reference acquired by multiview confocal super-resolution microscopy~\cite{wu2021multiview} (Fig.~\ref{fig6}c). An expanded mitochondrion was imaged, with its outer membrane and DNA immunolabeled by two distinct dyes. Notably, axial and lateral data distributions, as well as those between channels, exhibited significant dissimilarity. Applying SSAI-3D to each channel individually with customized sparse fine-tuning, we found that only SSAI-3D achieved high-fidelity restoration compared to existing methods (please refer to Supplementary Fig.~\ref{fig_Multiview} for method comparison). Analysis of the detected DNA volume within the image stack revealed that volume statistics from the SSAI-3D reconstruction were highly consistent with those derived from the near-isotropic reference (Fig.~\ref{fig6}d).

\section*{Discussion}
SSAI-3D demonstrated robust isotropic resolution recovery across diverse real-world microscopy systems and samples (Fig. \ref{fig5}). In particular, this approach was applied to label-free nonlinear microscopy of intact, thick, and heterogeneous tissues, where assumptions about data distribution (e.g., sparsity, label-specific distribution, axial-lateral similarity) and system priors (e.g., i.i.d. noise, LSI PSFs) often fail. The axial deblurring pipeline was further validated using publicly available, ground-truth-paired experimental datasets featuring potentially high lateral-axial dissimilarity in biological tissues, as well as unknown blurring and noise characteristics across different microscopy systems (Fig. \ref{fig6}). SSAI-3D achieved this robust axial deblurring by generating a PSF-varying, noise-resilient, sample-informed training dataset and sparsely fine-tuning a large pre-trained blind deblurring network.
Compared to existing computational methods, SSAI-3D does not require strong assumptions about the imaging system or sample data distribution, while maintaining low computational cost. Complementary to hardware-corrected isotropic imaging, SSAI-3D avoids the need for additional complex instrumentation and sample pre-processing, thus enabling broader applicability across biomedical applications, particularly for thick, intact, and living biosystems.

Despite the robustness, the performance of SSAI-3D will likely degrade in these scenarios. First, if the original axial resolution is more than 7 times worse than the lateral resolution (roughly 2.5\,\textmu m lateral resolution assuming a Gaussian beam), the recovery fidelity deteriorates gradually, as the ablation experiment shown in Fig. \ref{fig2}i, j. Secondly, the denoising step has been critical for high-fidelity axial deblurring, likely similar to the noise amplification in the classical deconvolution problem. The use of SSAI-3D should be cautionary when it comes to highly noisy datasets, where either more acquisition time or a more effective denoising algorithm could help.

\section*{Methods} \label{sec:methods}
\subsection*{Self-supervised dataset generation}
SSAI-3D employed a self-supervision strategy for learning isotropic restoration using the raw 3D anisotropic data itself. The self-supervised dataset was generated from the high-resolution lateral images, which were first denoised and then artificially blurred by convolving with 25 PSFs that combined five different sizes and five different orientations. 
For the denoising step, we adopted the framework presented in our previous work~\cite{ye2023learned}. To simulate an array of PSFs with varying sizes, Gaussian blur was applied to the lateral images using kernels with standard deviations ranging from 3 to 7 pixels. To simulate an array of PSFs with different orientations, the kernels were rotated within the range of $[-45^\circ, 45^\circ]$.

\subsection*{Contribution of different layers in fine-tuning}
To determine which layers to fine-tune for adapting to the data distribution of a customized dataset, we assessed each layer's contribution to the performance of the network using zero-shot metrics.
In this paper, we employed the following 14 such metrics, obtainable via a single forward and backward pass of the pre-trained network, to predict the performance of fine-tuning each specific layer.

We denote the weights of the pre-trained network as $\bm{\Theta}$ and the loss function as $\mathcal{L}$.
For layer $l$, we denote the single pass forward activation and the single pass backward gradient at this layer as $\bm{A}^{(l)}$ and $\bm{G}^{(l)}$, respectively.
Then, the 14 zero-shot metrics $\bm{U}^{(l)}=[u_1^{(l)}, ..., u_{14}^{(l)}]$ of layer $l$ can be represented as follows.
\begin{itemize}
    \item \textbf{Metrics evaluating forward activation.} These metrics aim to capture the compatibility of pre-trained parameters with the customized dataset. The metrics relate to activation values during the forward pass and include the average, standard deviation, average of absolute values, and standard deviation of absolute activation values:
    $$
    u_1^{(l)} = \mathbb{E}\left\{\bm{A}^{(l)}\right\}, u_2^{(l)} = \text{std}\left\{\bm{A}^{(l)}\right\}, u_3^{(l)}= \mathbb{E}\left\{\left\vert\bm{A}^{(l)}\right\vert\right\}, 
    u_4^{(l)} = \text{std}\left\{\left\vert\bm{A}^{(l)}\right\vert\right\}.
    $$
    \item \textbf{Metrics evaluating backward gradient.} These metrics aim to determine which pre-trained model parameters are most in need of adjustment. The metrics relate to the gradient of the model and include the average, standard deviation, average of absolute values, and standard deviation of absolute gradient values:
    $$
    u_5^{(l)} = \mathbb{E}\left\{\bm{G}^{(l)}\right\}, u_6^{(l)} = \text{std}\left\{\bm{G}^{(l)}\right\}, u_7^{(l)}= \mathbb{E}\left\{\left\vert\bm{G}^{(l)}\right\vert\right\}, 
    u_8^{(l)} = \text{std}\left\{\left\vert\bm{G}^{(l)}\right\vert\right\}.
    $$
    To evaluate the importance of each layer based on channel saliency, we also employed the $\mathtt{fisher}$ metric \cite{theis2018faster}. This metric assesses the impact of removing activation channels (and their corresponding parameters) that are estimated to have the least effect on the loss:
    $$
    u_9^{(l)} = \left(\bm{A}^{(l)} \odot \bm{G}^{(l)}\right)^2.
    $$
    \item \textbf{Metrics related to the loss function.} To evaluate the importance of each layer based on their sensitivity to the loss with respect to the network inputs, we employed the metric $\mathtt{snip}$ \cite{lee2018snip} as
    $$
    u_{10}^{(l)} =\left\vert{\frac{\partial \mathcal{L}}{\partial \bm\Theta} \odot \bm\Theta}\right\vert.
    $$
    To evaluate the importance of each layer based on their sensitivity to the gradient (as opposed to loss in $\mathtt{snip}$) with respect to the network inputs, we employed the metric $\mathtt{grasp}$ \cite{wang2020picking} as
    $$
    u_{11}^{(l)} =-\mathbb H\left\{\frac{\partial \mathcal{L}}{\partial \bm\Theta}\right\} \odot \bm\Theta,
    $$
    where $\mathbb H\{\cdot\}$ denotes the Hessian matrix.
    To evaluate the importance of each layer based on a loss function derived straightforwardly from the product of all the layer parameters (no need for input data to compute), we employed $\mathtt{synflow}$ \cite{tanaka2020pruning} as
    $$
     u_{12}^{(l)}= \frac{\partial \mathcal{L}}{\partial \bm\Theta} \odot \bm\Theta.
    $$
    \item \textbf{Metrics related to the pre-trained network.} We also included the metrics aim to capture the inherent knowledge from the parameters of the pre-trained network, including the average and the standard deviation:
    $$
    u_{13}^{(l)}= \mathbb{E}\left\{\bm{\Theta}^{(l)}\right\}, u_{14}^{(l)} = \text{std}\left\{\bm{\Theta}^{(l)}\right\}.
    $$
\end{itemize}

\subsection*{Sparse fine-tuning with a surgeon network}
Sparse fine-tuning was achieved by selecting specific layers in the large pre-trained network for modification while freezing the remaining layers. To efficiently select these layers based on the aforementioned zero-shot metrics, we employed a small surgeon network.
The surgeon network was a 6-layer multi-layer perceptron (MLP) with 200k parameters (Supplementary Fig. \ref{fig_network}). This network took the zero-shot metrics $\bm U^{(l)}$, along with the positional embedding $l$, as inputs and outputs a score $s^{(l)}$, representing the contribution of layer $l$ in the pre-trained deblurring network.
The layers with the highest 10\% scores were then selected for sparse fine-tuning.
To pre-train the surgeon network using ground truth labels, we created a meta-dataset ($\sim$1\,GB) comprising 5 different stacks from a custom-built nonlinear microscope~\cite{liu2024deep}.
Using this meta-dataset, 300 input-output data pairs were generated, following the format $\left\{\left[\bm U^{(l)},l\right], s^{(l)}\right\}$ as defined above.
In training the surgeon network, we employed the Adam optimizer with a learning rate of $10^{-4}$, a weight decay of $10^{-6}$, and the pairwise ranking loss, as we were only interested in the relative performance of the layers.
The pre-trained surgeon network is released along with the codes.

\subsection*{Architecture, training, and inference of deblurring network}
The deblurring network employed a NAFNet topology with an encoder-decoder architecture for end-to-end deconvolution~\cite{chen2022simple}. NAFNet is a pre-trained network on real-world deblurring tasks, such as those in the GoPro~\cite{park2019down} and REDS~\cite{liu2022video} datasets.
Due to the inherent nature of the method, the network was implemented in 2D, with input patches of 256$\times$256 and a batch size of 8. For all presented experiments, network parameters were optimized using the Adam optimizer with a learning rate of $10^{-3}$ and weight decay of $10^{-4}$. Training SSAI-3D took approximately 30 minutes on a single Nvidia GeForce RTX 3080 card, while inference for a 300$\times$300$\times$300 tensor took roughly 50 seconds. To minimize memory consumption, the model was quantized to float16 precision.
The employed deblurring network contained approximately 100 million parameters. Only the deconvolution network was trained, with $\sim$10\% of the layers fine-tuned and the remaining $\sim$90\% frozen. In the inference stage, all layers were fixed, and axial images were first denoised using the same denosing network and then deconvolved to maintain high data resemblance with the training data.

\subsection*{Simulation data generation}
For the simulation of beads in Fig.~\ref{fig2}a, the ground truth image volumes were generated with 50 randomly located beads of random radius and intensity. The isotropic beads were generated according to a 3D Gaussian distribution with fixed standard deviations of 2.0 in all three dimensions. The volume for training the model contained 50 beads in a 30\,\textmu m$\times$30\,\textmu m$\times$30\,\textmu m volume, with 100\,nm pixel size.
For the simulation of symmetric tubular strands in Fig. \ref{fig4}c, the ground truth image volumes contained strands facing random directions at random locations, with a mean thickness of 4 pixels and intensity values ranging from 15000 to 65535. The volume for training the model contained 5000 tubes in a 300\,\textmu m$\times$300\,\textmu m$\times$300\,\textmu m volume. A 3D elastic grid-based deformation field was then applied to deform the volume, creating the ground truth. Different blur levels were generated using a 51$\times$51$\times$51 3D Gaussian kernel, elongated axially to serve as raw input, with a standard deviation of 0.5 in lateral directions and 4.0 in the axial direction.
For the simulation of asymmetric tubular strands in Fig. \ref{fig2}b, the ground truth volume was generated via 3D Bézier curves, shaped as curves in the lateral direction and straight horizontal lines axially. This created semi-synthetic data with vastly different distributions across views. A total of 5000 strands were generated in a 300\,\textmu m$\times$300\,\textmu m$\times$300\,\textmu m volume, with a thickness of 1 to 4 pixels, a total length of around 1200 pixels, and binary brightness values. Similar to the previous simulation, blur was introduced using a 51$\times$51$\times$51 3D Gaussian kernel with anisotropic standard deviations in lateral and axial directions.

\subsection*{Sample preparation and imaging acquisition}
3D imaging in living tissues was achieved using a custom-built nonlinear microscope, as detailed in~\cite{qiu2024spectral}. The preparation of the living and intact human blood-brain barrier microfluidic model (used in Fig.~\ref{fig3}e) was described in~\cite{liu2024deep} and the preparation of freshly excised human endometrial tissue (used in Fig.~\ref{fig5}b and f) was described in~\cite{ye2023learned}. A custom-built multimode fiber source at 1100\,nm was employed for simultaneous label-free imaging of THG, NAD(P)H, SHG, and FAD~\cite{you2018intravital,liu2024deep}. To achieve deep imaging with uniform SNR across different depths, pulse energy was varied according to imaging depth, maintaining approximately 3\,nJ pulse energy at the excitation focus.

\subsection*{Method comparison}
We compared the performance of SSAI-3D with four baseline methods: Richardson-Lucy, OT-CycleGAN \cite{park2022deep}, Self-Net \cite{ning2023deep}, and CARE \cite{weigert2018content}. These methods were all implemented by open-source codes released by the cited papers. 
For CARE, the semi-synthetic dataset was created by applying an explicit blurring and down-sampling model to degrade the high-resolution lateral images to resemble the low-resolution axial images.
In Fig.~\ref{fig2}i and j, a Gaussian blur with a standard deviation of 5 was assumed to be known for CARE. To assess robustness against PSF mismatch, the model trained with this known blur was applied to different levels of blurring for CARE, while a single model trained on varying blurs was used for SSAI-3D. 
In Fig.~\ref{fig6}a and c, where the blurring was spatially varying, a single blur within the range of the axial degradation was selected.
In Fig.~\ref{fig3}b, Fig.~\ref{fig3}e, and Supplementary Fig.~\ref{fig_finetuning}, SSAI-3D with full fine-tuning was implemented by fine-tuning the entire pre-trained deblurring network without freezing any layers.
For all the experiments presented, the hyper-parameters and training settings were universal.

\subsection*{Evaluation metrics}
The MSE used in the paper is defined as 
$$
\text{MSE} = \frac{1}{N}\Vert \bm X-\bm X_0 \Vert^2,
$$
where $\bm X$ denotes the restored image, $\bm X_0$ denotes the corresponding ground truth, and $N$ denotes the number of pixels. 
The SSIM used in the paper is defined as
$$
\text{SSIM} = \frac{(2\mu\mu_0+C_1)(2\sigma'+C_2)}{(\mu^2+\mu_0^2+C1)(\sigma^2+\sigma_0^2+C_2)},
$$
where $\{\mu,\mu_0\}$ and $\{\sigma,\sigma_0\}$ are the means and variances of $\bm X$ and $\bm X_0$, respectively.
$\sigma^\prime$ is the covariance of $\bm X$ and $\bm X_0$. The two constants $C_1=(k_1L)^2$ and $C_2=(k_2L)^2$ are chosen with $k_1=0.01$ and $k_2=0.03$.
In Fig. \ref{fig2}c, the axial resolution after isotropic recovery was measured by the full width at half maximum (FWHM) of the sub-diffraction-limited beads in $z$-direction (mean$\pm$standard deviation, $n=50$). In Fig. \ref{fig5}, the resolution of the image was characterized by the width in its Fourier spectrum. In Fig. \ref{fig6}b and d, the individual granules were detected by an universal threshold of pixel value in raw, software-corrected, and hardware-corrected images.
The detection accuracy of image $\bm X$ was characterized by the normalized cross-correlation (NCC) against hardware-corrected image $\bar{\bm X}$ as
$$
\text{NCC} = \frac{\left(\bm X-\mathbb E \{\bm X\}\right)\odot\left(\bar{\bm X}-\mathbb E \{\bar{\bm X}\}\right)}{\sqrt{\left\Vert\bm X-\mathbb E \{\bm X\}\right\Vert^2 \left\Vert\bar{\bm X}-\mathbb E \{\bar{\bm X}\}\right\Vert^2}}.
$$

\section*{Data availability}
The open-source datasets used in this paper can be found at \href{https://doi.org/10.5281/zenodo.7882519}{https://doi.org/10.5281/zenodo.7882519} \cite{ning2023deep} (Fig. \ref{fig5}a, c, and e), \href{https://idr.openmicroscopy.org/}{https://idr.openmicroscopy.org/} \cite{voigt2019mesospim} (under accession number idr0066, Fig. \ref{fig6}a), and \href{https://zenodo.org/record/5495955}{https://zenodo.org/record/5495955} \cite{wu2021multiview} (Fig. \ref{fig6}b).
Other datasets are available from the corresponding author upon reasonable request.

\section*{Code availability}
The implementations of SSAI-3D, the pre-trained surgeon network, as well as some examples
will be publicly available.

\bibliographystyle{naturemag}

\bibliography{ISO}

\section*{Acknowledgements}
We thank Honghao Cao and Tong Qiu for assistance in nonlinear imaging, as well as helpful discussions from Matthew Yeung, Steven F. Nagle, and Haidong Feng.
We thank Song Han for insights into sparse fine tuning.
We thank Roger D. Kamm, Sarah Spitz, and Francesca Michela Pramotton for providing the human blood-brain barrier microfludic model for nonlinear imaging.
We thank Ellen L. Kan for assistance in preparing the human endometrium tissue for nonlinear imaging.
We thank Fan Wang, Manuel Levy, and Eva Lendaro for providing the mouse whisker pad tissue for nonlinear imaging.
We acknowledge that Fig. \ref{fig1}a contains materials from BioRender (https://biorender.com/). The work has been supported by MIT startup funds and CZI Dynamic Imaging via Chan Zuckerberg Donor Advised Fund (DAF) through the Silicon Valley Community Foundation (SVCF).
K.L. acknowledges support from the MIT Irwin Mark Jacobs (1957) and Joan Klein Jacobs Presidential Fellowship.

\section*{Author contributions}
J.H., K.L., and S.Y. conceived the idea of the project.
J.H. designed detailed implementations in network training and inference.
K.L. built the optical setup and performed the imaging experiments.
J.H. and K.L. analyzed the data and prepared the figures with input from K.B.I., K.M., and L.G.G..
K.B.I. and L.G.G. provided the samples for imaging and insights into the biomedical questions.
J.H., K.L., and S.Y. wrote the manuscript with the input from all authors.
S.Y. obtained the funding and supervised the research.

\section*{Competing interests}
The authors declare no competing interests.

\newpage
\section*{Supplementary information}

\renewcommand{\thefigure}{S\arabic{figure}}
\setcounter{figure}{0}

\begin{figure}[htbp]
\centering
\includegraphics[width=\textwidth]{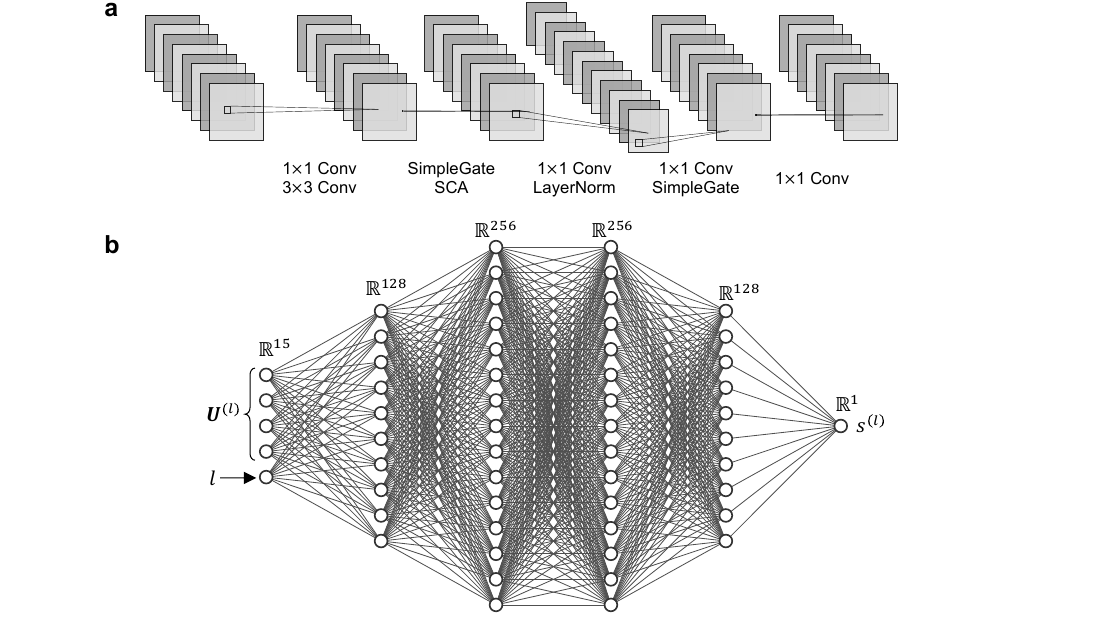}
\caption{Network architecture of the deblurring network (intra-block structure, \textbf{a}) and the surgeon network (\textbf{b}).
The deblurring network is a large convolutional neural network (NAFNet \cite{chen2022simple}).
The surgeon network is a small multi-layer perceptron (MLP) that takes the zero-shot metrics $\bm U^{(l)}$ and positional embedding $l$ and outputs a score $s^{(l)}$. 
}
\label{fig_network}
\end{figure}

\vspace{10mm}

\begin{figure}[htbp]
\centering
\includegraphics[width=\textwidth]{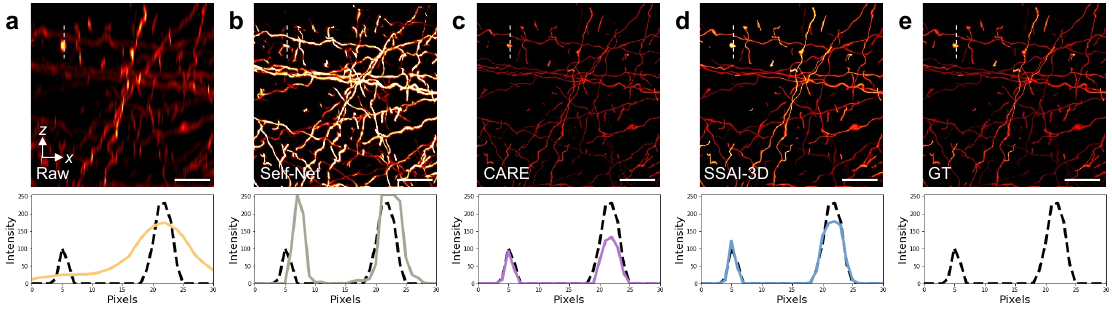}
\caption{Comparison of co-registered intensity profiles in the ground truth image (\textbf{e}) with the raw axial image (\textbf{a}) as well as the resolution recovered image using Self-Net (\textbf{b}), CARE (\textbf{c}), and 
SSAI-3D (\textbf{d}). 
Scale bars: 60\,\textmu m.}
\label{fig_line}
\end{figure}

\begin{figure}[htbp]
\centering
\includegraphics[width=\textwidth]{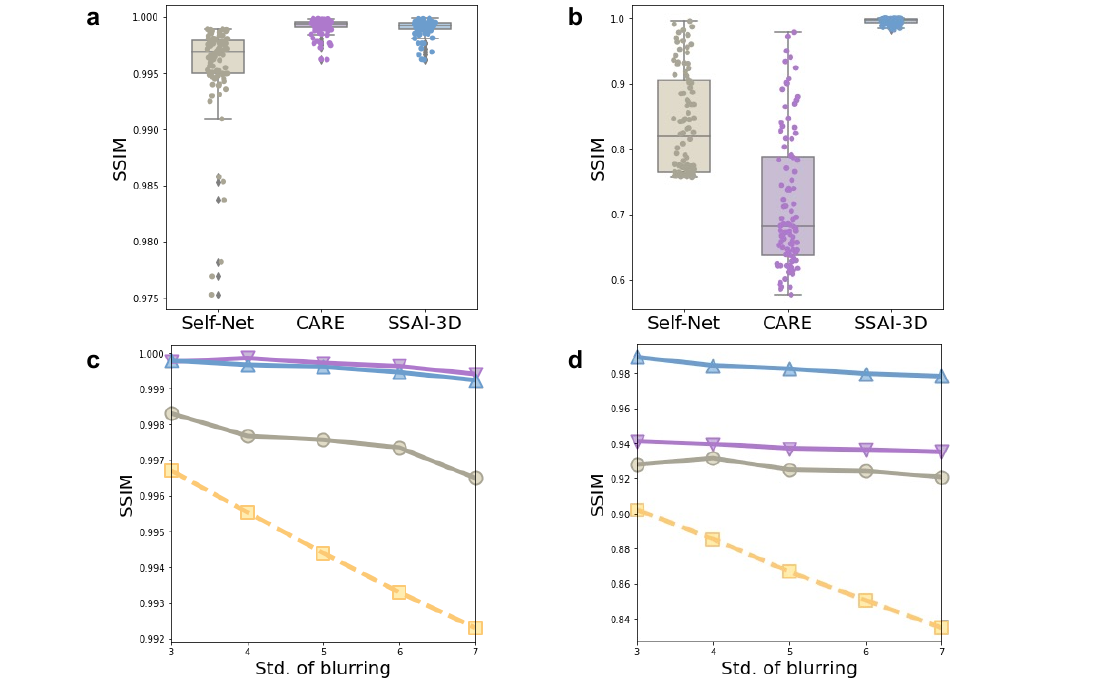}
\caption{Fidelity of resolution recovery characterized by SSIM.
\textbf{a, b}, SSIM of resolution recovery in Fig. \ref{fig2}c and d.
\textbf{c, d}, Ablation study on the size of PSF for resolution recovery in Fig. \ref{fig2}c and d.}
\label{fig_SSIM}
\end{figure}

\begin{figure}[htbp]
\centering
\includegraphics[width=\textwidth]{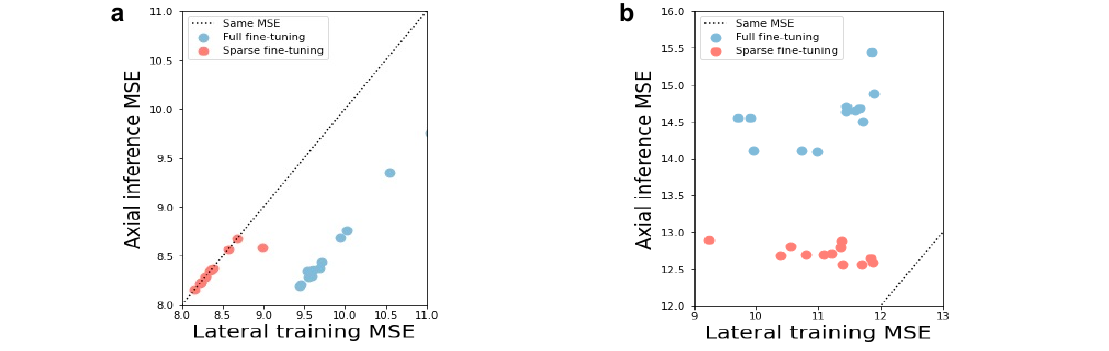}
\caption{Comparison of full fine-tuning and sparse fine-tuning in lateral training MSE against axial inference MSE in the same number of training epochs. \textbf{a} and \textbf{b} correspond to the simulation of spheres and cylinders in Fig. \ref{fig3}a and b, respectively.}
\label{fig_finetuning}
\end{figure}

\begin{figure}[htbp]
\centering
\includegraphics[width=\textwidth]{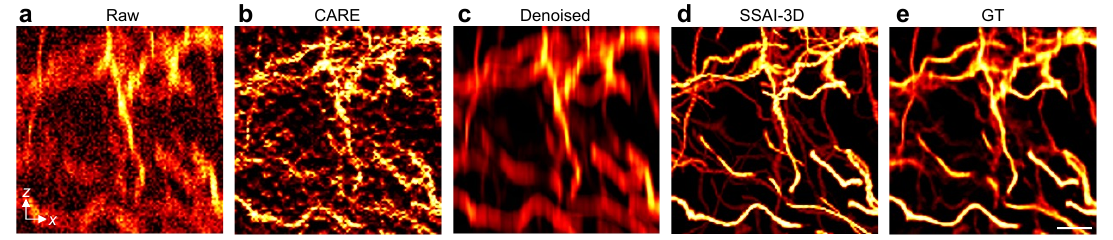}
\caption{Denoising in SSAI-3D improves the robustness of deblurring against noise.
\textbf{a}, Simulation of 3D imaging stack with noise.
\textbf{b}, Deblurring result of CARE.
\textbf{c}, Denoised image of \textbf{a}.
\textbf{d}, Deblurring result of SSAI-3D after denoising.
\textbf{e}, Ground truth image.
Scale bars: 10\,\textmu m. }
\label{figS_denoise}
\end{figure}

\begin{figure}[htbp]
\centering
\includegraphics[width=\textwidth]{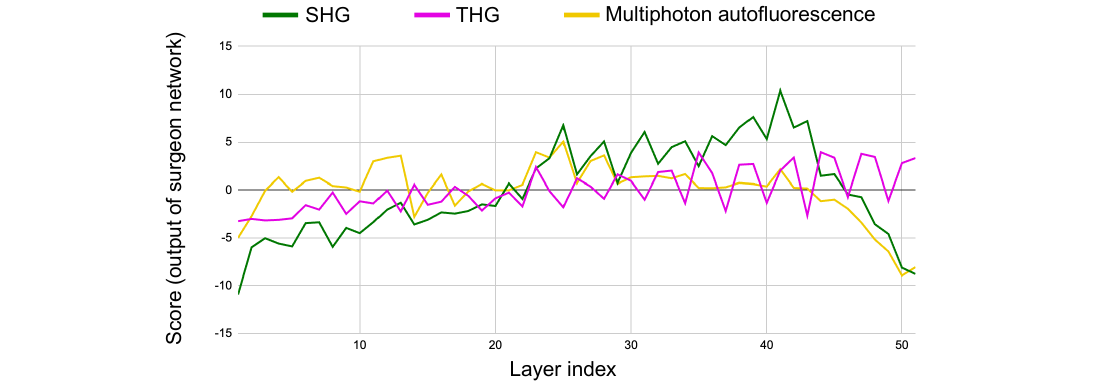}
\caption{Contribution analysis shows the impact of fine-tuning different layers in the deblurring network. Different imaging stacks have different preferences on the layers to fine-tune, evidenced by the scores (output $s^{(l)}$ of the surgeon network). SHG, THG, and multiphoton autofluorescence correspond to images presented in Fig. \ref{fig5}b, d, and f, respectively. }
\label{figS_contribution}
\end{figure}

\begin{figure}[htbp]
\centering
\includegraphics[width=\textwidth]{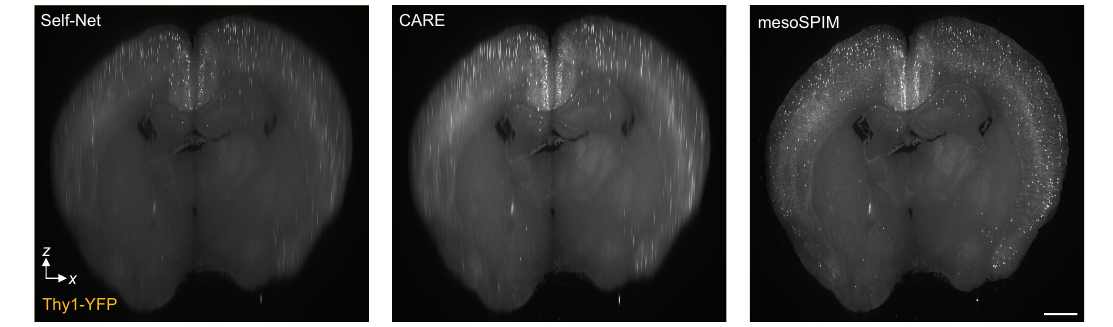}
\caption{Comparison of existing methods on the mesoSPIM dataset \cite{voigt2019mesospim}. SSAI-3D exhibited higher fidelity in reconstruction and higher accuracy in downstream analysis (Neuron extraction in Fig. \ref{fig6}a and b). Scale bars: 1\,mm. }
\label{fig_ASLM}
\end{figure}

\begin{figure}[htbp]
\centering
\includegraphics[width=\textwidth]{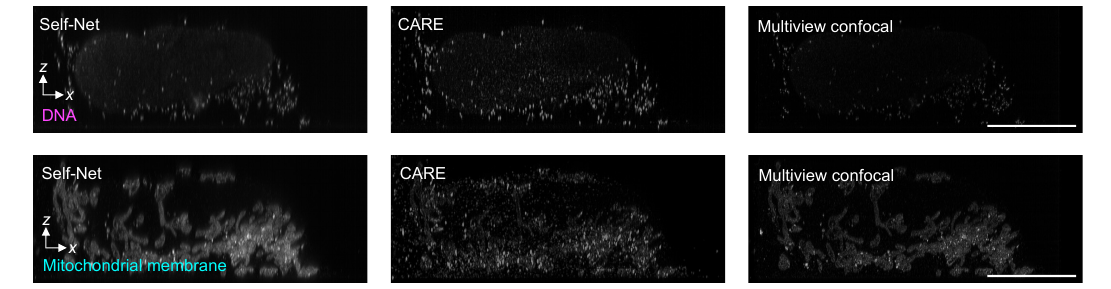}
\caption{Comparison of existing methods on the multiview confocal dataset \cite{wu2021multiview}. SSAI-3D exhibited higher fidelity in reconstruction and higher accuracy in downstream analysis (Statistics of DNA volume in Fig. \ref{fig6}c and d). Scale bars: 5\,\textmu m.}
\label{fig_Multiview}
\end{figure}

\renewcommand{\thetable}{S\arabic{table}}
\setcounter{table}{0}
\begin{table}[]
\caption{Summary of 3D imaging data used in this work.}\label{sum}
\label{summary_table}
\centering
\begin{tabular}{@{}c|cccc@{}}
\toprule
Figure & System      & Sample                                                                                        & \begin{tabular}[c]{@{}c@{}}Pixel number\\ ($X,Y,Z$)\end{tabular} & \begin{tabular}[c]{@{}c@{}}Pixel size\\ ($\text{d}x,\text{d}y,\text{d}z$)\,\textmu m\end{tabular} \\ \midrule
\ref{fig2}a     & Simulation  & Symmetric beads                                                                               & (300, 300, 300)                                                  & (0.1, 0.1, 0.1)                                                                 \\
\ref{fig2}b     & Simulation  & Asymmetric strands                                                                            &  (900, 900, 900)                                                                &    (0.33, 0.33, 0.33)                                                                                \\
\ref{fig3}a     & Simulation  & Symmetric spheres                                                                             &  (300, 300, 300)                                                                &           (1.0, 1.0, 1.0)                                                                        \\
\ref{fig3}b     & Simulation  & Asymmetric cylinders                                                                          &   (300, 300, 300)                                                               &            (1.0, 1.0, 1.0)                                                                      \\
\ref{fig3}e     & Multiphoton & \begin{tabular}[c]{@{}c@{}}Living and intact\\ human blood-brain\\ barrier model \cite{liu2024deep}\end{tabular} &  (700, 700, 250)                                                                &                                 (0.5, 0.5, 2.0)                                                   \\
\ref{fig3}f     & Nonlinear   & \begin{tabular}[c]{@{}c@{}}Freshly excise human\\ endometrium tissue\end{tabular}             &  (512, 512, 60)                                                                &     (0.4, 0.4, 1.2)                                                                             \\
\ref{fig4}      & Simulation  & Symmetric strands                                                                             &      (300, 300, 300)                                                            &                                       (1.0, 1.0, 1.0)                                             \\
\ref{fig5}a     & Light sheet & \begin{tabular}[c]{@{}c@{}}Cleared and stained\\mouse brain \\vasculature \cite{todorov2020machine}  \end{tabular}                                                                            &                                                     (916, 916, 498)             &                                      (1.63, 1.63, 3)                                        \\
\ref{fig5}b     & Multiphoton & \begin{tabular}[c]{@{}c@{}}Freshly excise human\\ endometrium tissue\end{tabular}             &(512, 512, 60)&(0.4, 0.4, 1.2)\\
\ref{fig5}c     & Confocal    & \begin{tabular}[c]{@{}c@{}}Cleared and stained \\mouse brain\\neurons \cite{ning2023deep}  \end{tabular}     &  (1024, 1024, 90)&       (0.21, 0.21, 1.00)\\
\ref{fig5}d     & SHG         & \begin{tabular}[c]{@{}c@{}}Fixed mouse whisker\\ pad tissue \cite{qiu2024spectral}\end{tabular}         & (500, 500, 50)&(0.5, 0.5, 1.0)\\
\ref{fig5}e     & Wide-field  & \begin{tabular}[c]{@{}c@{}}Cleared\\ mouse liver \cite{ning2023deep}\end{tabular}&(1800, 1700, 246)&(0.32, 0.32, 1.00)\\
\ref{fig5}f     & THG         & \begin{tabular}[c]{@{}c@{}}Freshly excise human\\ endometrium tissue\end{tabular}              &(512, 512, 60)&(0.4, 0.4, 1.2)\\
\ref{fig6}a     & Light sheet & \begin{tabular}[c]{@{}c@{}}Cleared and stained\\ mouse brain \cite{voigt2019mesospim}\end{tabular}                                                                        &           (2048, 2048, 1937)&(6.4, 6.4, 4.8)\\
\ref{fig6}c     & Confocal    & \begin{tabular}[c]{@{}c@{}}Expanded\\ mitochondria \cite{wu2021multiview}\end{tabular}                                                              &      (530, 1396, 1396)&(0.013, 0.013, 0.016)\\ \bottomrule
\end{tabular}
\end{table}
\end{document}